\documentclass[aps,prd,twocolumn,showpacs,unsortedaddress,superscriptaddress,showkeys,preprintnumbers,letterpaper]{revtex4-1}
\pdfoutput=1

\usepackage{acronym}
\usepackage{amsmath}
\usepackage{amssymb}
\usepackage{graphicx}
\usepackage{subfigure}
\usepackage[usenames]{color}
\usepackage[normalem]{ulem}
\usepackage[colorlinks]{hyperref} 
\newcommand{\mat}[1]{{\bf #1}}

\newcommand*{\diff}{\,\mathrm{d}}
\newcommand{\T}{{\mathrm{T}}}

\begin{document}

\newcommand{\LIGOCaltech}{LIGO Laboratory, California Institute of Technology, 
Pasadena, CA 91125, USA}
\newcommand{\TAPIR}{Theoretical Astrophysics, California Institute of
Technology, Pasadena, CA 91125, USA}
\newcommand{\CITA}{Canadian Institute for Theoretical Astrophysics, 60 St.
George Street, University of Toronto, Toronto, ON M5S 3H8, Canada}
\newcommand{\Perimeter}{Perimeter Institute for Theoretical Physics, Waterloo,
Ontario N2L 2Y5, Canada}
\newcommand{\AEI}{Albert-Einstein-Institut, Max-Planck-Institut f\"ur
Gravitationsphysik, D-30167 Hannover, Germany}
\newcommand{\Leibniz}{Leibniz Universit\"at Hannover, D-30167 Hannover, Germany}

\newcommand{\thetabar}{\bar{\theta}}

\title{
Composite gravitational-wave detection of compact binary coalescence
}

\date{\today}

\author{Kipp~Cannon}
\email{kipp.cannon@ligo.org}
\affiliation{\LIGOCaltech}
\affiliation{\CITA}

\author{Chad~Hanna}  
\email{chad.hanna@ligo.org}
\affiliation{\LIGOCaltech}
\affiliation{\Perimeter}

\author{Drew~Keppel}  
\email{drew.keppel@ligo.org}
\affiliation{\LIGOCaltech}
\affiliation{\TAPIR}
\affiliation{\AEI}
\affiliation{\Leibniz}

\author{Antony~C.~Searle}
\email{antony.searle@ligo.org}
\affiliation{\LIGOCaltech}


\begin{abstract}
The detection of gravitational waves from compact binaries relies on a
computationally burdensome processing of gravitational-wave detector data. The
parameter space of compact-binary-coalescence gravitational waves is large and
optimal detection strategies often require nearly redundant calculations.
Previously, it has been shown that singular value decomposition of search
filters removes redundancy.  Here we will demonstrate the use of singular value
decomposition for a composite detection statistic.  This can greatly improve
the prospects for a computationally feasible rapid detection scheme across a
large compact binary parameter space.
\end{abstract}



\preprint{}

\maketitle

\acrodef{BNS}{binary neutron star}
\acrodef{CBC}{compact binary coalescence}
\acrodef{GW}{gravitational-wave}
\acrodef{PN}{post-Newtonian}
\acrodef{ROC}{Receiver Operator Characteristic}
\acrodef{SNR}{signal-to-noise ratio}
\acrodef{SPA}{stationary phase approximation}
\acrodef{SVD}{singular value decomposition}


\section{Introduction}

Ground-based laser-interferometric gravitational-wave detectors have
demonstrated sensitivity to gravitational-wave strain at the level of
$10^{-22}$ or better over a band from $80-1000$ Hz~\cite{LIGO2009, Virgo2008,
GEO2010}.  Later this decade advanced detectors will surpass the present
sensitivity by a factor of $\sim 10$~\cite{advLIGO, advVirgo2005}.  One of the
most promising sources of gravitational waves is the merger of two compact
objects~\cite{LastThree}.  Advanced detectors are expected to detect on the
order of 40 neutron star - neutron star merger events per
year~\cite{kopparapu2008, rates2010} and a similar number of mergers involving
black holes.  The exact parameters of the signals, such as mass and spin of the
component objects, will not be known ahead of time.  Therefore, the optimal
detection strategy must include the possibility of detecting signals with
unknown parameters.  Matched filtering has been employed to search this
parameter space~\cite{finn1993, findchirppaper}.  The parameter space is
explored by choosing a discrete set of filters that guarantees that all signals
within the parameter space are found with a \ac{SNR} greater than
$\sim97\%$~\cite{Owen:1995tm} of the maximum possible.   Recently it has been
shown that the use of singular value decomposition can reduce the number of
filters necessary to search the parameter space~\cite{SVDpaper}. This paper
extends that work to explore use of the \ac{SVD} filter outputs for detection
without reconstructing the physical template waveforms.

If one is interested only in knowing whether any of the signals are present,
and not which one is present, then the problem of detection decouples from that
of parameter estimation.  Wainstein and Zubakov (1962)~\cite{wainstein:1962}
describe the problem of detecting any of several signals without concern for
parameter estimation as composite detection.  This paper will explore the
composite detection of compact binary signals using the techniques proposed
in~\cite{SVDpaper}.  We find that the composite detection statistics explored
produce a lower detection efficiency at a fixed false alarm rate than
traditional approaches.  However when combined hierarchically with traditional
approaches, the combined approach can perform as well at low false alarm rate,
but with reduced computational cost.


\section{Composite detection of compact binary signals}

We will consider an unknown gravitational wave signal arising from a compact
binary coalescence in the digitized gravitational-wave detector output as a
vector of data points parameterized by an unknown amplitude $A$ and a
collection of unknown parameters $\thetabar$ (e.g., the masses of the component
bodies) that determine the shape of the signal.  We will denote this signal as
$A \mat{s}(\thetabar)$.  Both the amplitude and shape parameters are not known
\emph{a priori}. The relative frequency of parameters in the population is
described by the probability density function $p(A, \thetabar)$.  We will
assume the joint distribution $p(A, \thetabar)$ is separable, that is $p(A,
\thetabar) = p(A)p(\thetabar)$.  This is not generally true globally across the
parameter space, but should be roughly true locally.  We will denote the vector
of discretely time-sampled strain data as $\mat{h}$.  In addition to containing
normally distributed noise, $\mat{n}$, $\mat{h}$ will possibly contain the
gravitational-wave signal,
%
%
\begin{equation}
\label{eq:det_output}
\mat{h} = \mat{n} + A \mat{s}(\thetabar).
\end{equation}
Assuming the noise has unit variance and $\mat{s}(\thetabar)$ is normalized
such that the inner-product of it with itself is unity, $A$ is the expected
value for the \ac{SNR} of a signal after optimally filtering.

The optimal detection statistic is the marginalized likelihood
%
%
\begin{equation}
\label{eqn:lambda}
\Lambda = \frac{p(\mat{h}|\mat{s})}{p(\mat{h}|\mat{0})} ,
\end{equation}
where $p(\mat{h}|\mat{s})$ is the probability of obtaining $\mat{h}$ given the
presence of any signal, with the signal parameters $A,\thetabar$ integrated
out
%
%
\begin{eqnarray}
\label{eq:unmarginalized_likelihood}
p(\mat{h}|\mat{s}) & = &
\int_V \int_{-\infty}^{+\infty} p(\mat{h}|A,\thetabar) p(\thetabar) p(A) \diff A \diff \thetabar ,
\end{eqnarray}
and $p(\mat{h}|0)$ is the probability of obtaining $\mat{h}$ in the absence of
any signal.  The marginalized likelihood increases as the probability of the
data containing a signal increases.  For white noise,
%
%
\begin{equation}
\label{eqn:pnoise}
p(\mat{h}|0) \propto \exp\left(-\frac{1}{2}\mat{h}^\T\mat{h}\right) .
\end{equation}
This white noise form is also valid for colored noise cases if one applies the
linear whitening transformation to the data $\mat{h}$.  In the frequency
domain, this transformation divides the data by the amplitude spectral density
of the noise.  Our goal is to find a computationally inexpensive approximation
to this optimal detection statistic using the \ac{SVD}-reduced filter set
described in~\cite{SVDpaper}.

\subsection{Expansion of the marginalized likelihood}
\label{subsec:Expansion}

In this section we will consider an expansion of the marginalized likelihood
and show that it gives rise to a detection statistic that exploits the SVD
filter basis.  

For simplicity we will assume that the signal parameters $\thetabar$ take on
discrete values~\footnote{In practice, discretizing the parameter space is
already done when choosing the initial filter bank~\cite{Owen:1995tm}.} so that
we may index different signals as $\bf{s}_i$.  We will also assume that the
amplitude, $A$, does not depend on the signal index $i$.  The detector output
now has the form
%
%
\begin{equation}
\label{eq:discrete_parameter_strain}
\mat{h} = \mat{n} + A \mat{s_i}.
\end{equation}
We don't know \emph{a priori} which signal $\mat{s_i}$ is present in the data.
The marginalized likelihood~\cite{finn1993} is
%
%
\begin{equation}
\label{eq:disc_marg_like}
\Lambda \propto \sum_j \exp[A \mat{h} \cdot \mat{s_j}],
\end{equation}
where the integral in \eqref{eq:unmarginalized_likelihood} is replaced by a sum
over all possible templates $\mat{s_j}$.  Expanding the exponential to second
order yields
%
%
\begin{equation}
\label{eq:exponentialexpansion}
\Lambda \stackrel{\sim}{\propto} \sum_j \left[ 1 + A \mat{h} \cdot \mat{s_j} + \frac{1}{2} A^2 (\mat{h} \cdot \mat{s_j})^2 \right] . 
\end{equation}
The first order term is oscillatory and contributes little to the sum.
We thus ignore it and define the approximate likelihood as $\Lambda'$
%
%
\begin{equation}
\label{eq:approxlikelihood}
\Lambda' := \sum_j (\mat{h} \cdot \mat{s_j})^2 . 
\end{equation}
Using the change of basis described in~\cite{SVDpaper}, where
%
%
\begin{eqnarray}
\label{eq:SVDbasis}
\mat{s_j} &=& \sum_{k} v_{jk} \sigma_k {\bf u_k},
\end{eqnarray}
the approximate likelihood expression \eqref{eq:approxlikelihood} simplifies to
%
%
\begin{eqnarray}
\label{eq:low_amp_likelihood}
\Lambda' &=& \sum_k (\sigma_k {\bf h} \cdot {\bf u_k})^2 ,
\end{eqnarray}
due to the properties of the orthonormal matrix $v_{jk}$.
\eqref{eq:low_amp_likelihood} has no dependence on the original parameter index
$j$.  As was shown in~\cite{SVDpaper}, fewer filters are required using the
$\mat{u_k}$ basis, therefore the sum over $k$ has fewer terms than the original
sum over $j$.  Unfortunately higher order terms in the expansion of
\eqref{eq:exponentialexpansion} do not simplify as well.  It is possible to
derive detection statistics that perform better than
\eqref{eq:low_amp_likelihood} while still exploiting the reduced filter set;
one such example is shown in the next section.

\subsection{Assuming the signal probability distribution is a multi-variate normal distribution}

In this section we explore approximating the signal probability
$p(\mat{h}|\mat{s})$ in \eqref{eq:unmarginalized_likelihood} as a multivariate
normal distribution.  This will result in a different detection statistic that
still exploits the SVD filter basis.

We begin by assuming the signal probability
\eqref{eq:unmarginalized_likelihood} has the form
%
%
\begin{equation}
\label{eqn:psignal}
p(\mat{h}|\mat{s}) \propto \exp\left(-\frac{1}{2}\mat{h}^\T\mat{C}^{-1}\mat{h}\right) ,
\end{equation}
whose covariance matrix $\mat{C}$ of second-order moments completely
defines it. The covariance matrix is
%
%
\begin{eqnarray}
\label{eqn:Cij}
C_{ij} &=& \mathrm{cov}(h_i,h_j) \nonumber\\
&=& \left\langle (n_i + A s_i(\thetabar) + )(n_j +A s_j(\thetabar))\right\rangle \nonumber\\
&=& \left\langle n_i n_j\right\rangle + \left\langle A^2\right\rangle \left\langle s_i(\thetabar) s_j(\thetabar)\right\rangle \nonumber\\
&=& \delta_{ij} +  \left\langle A^2\right\rangle \int_V s_i(\thetabar) s_j(\theta) p(\thetabar)\,\mathrm{d}\thetabar ,
\end{eqnarray}
where $\left\langle A^2\right\rangle = \int_{-\infty}^{+\infty}A^2p(A)\,\diff
A$ and we have assumed the independence of the variables $A$, $\thetabar$ and
$n_i$.

Using \eqref{eqn:pnoise} and \eqref{eqn:psignal}, we compute the logarithm of
the marginalized likelihood, $\Lambda$. This is monotonic in $\Lambda$ and
therefore is sufficient to use as a ranking statistic
%
%
\begin{equation}
\label{eqn:approxlambda}
\ln\Lambda \propto \Gamma := \mat{h}^\T \left(\mat{I} - \mat{C}^{-1}\right) \mat{h} .
\end{equation}
The performance of \eqref{eqn:approxlambda} depends on the waveform population;
we will assess it for simple but realistic cases in Sec.~\ref{sec:ROC}.

We compute $\mat{C}$ by drawing $N$ samples of $\thetabar$ distributed
according to $p(\thetabar)$. We assume a signal population that is distributed
according to our ability to distinguish signals, which is approximated by the
filter banks described in~\cite{Owen:1995tm}.  This definition of the signal
population is approximately uniform in the $\thetabar$ distribution for
sufficiently small regions of parameter space. The integral in \eqref{eqn:Cij}
is then approximated by a summation over these samples,
%
%
\begin{gather}
C_{ij} = \delta_{ij} + \frac{\left\langle A^2\right\rangle}{N} \sum^N_{k=1} s_i(\theta_k) s_j(\theta_k) \nonumber \\
\mat{C} = \mat{I} + \frac{\left\langle A^2\right\rangle}{N} \sum_{k=1}^N \mat{s}^\T(\theta_k) \mat{s}(\theta_k) ,
\end{gather}
where $\theta_k$ are discrete values of $\thetabar$ and $\mat{s}(\theta_k)$ is
a row vector of the time samples $s_i(\theta_k)$.  By defining the matrix
$S_{ik} = s_i(\theta_k)$ we can simplify the notation to
%
%
\begin{equation}
\mat{C} = \mat{I} + \frac{\left\langle A^2\right\rangle}{N} \mat{S}^\T \mat{S} .
\end{equation}
This is precisely the arrangement of the signal matrix proposed
in~\cite{SVDpaper}.
As in \eqref{eq:SVDbasis}, we use the \ac{SVD} to decompose the signal matrix
$\mat{S}$ (here written in matrix notation but equivalent to
\eqref{eq:SVDbasis}),
%
%
\begin{equation}
\mat{S} = \mat{V} \mat{\Sigma} \mat{U}^\T,
\end{equation}
where $\mat{V}$ and $\mat{U}$ are unitary matrices and $\mat{\Sigma}$ is a
diagonal matrix of the singular values of $\mat{S}$, referred to by
components ${\sigma_k}$ in the Sec. \ref{subsec:Expansion}.  $\mat{U}$ is the
matrix of the orthonormal basis vectors $\mat{u_k}$ described in the previous
section.  Then we note that
%
%
\begin{align}
\mat{I} - \mat{C}^{-1} &= \mat{I} - \left(\mat{I} + \frac{\left\langle A^2\right\rangle}{N} \mat{S}^\T\mat{S}\right)^{-1} \nonumber\\
&= \mat{I} - \left(\mat{I} + \frac{\left\langle A^2\right\rangle}{N} \mat{U} \mat{\Sigma}^2 \mat{U}^\T\right)^{-1} \nonumber\\
&= \mat{U} \left(\mat{I} - \left(\mat{I} + \frac{\left\langle A^2\right\rangle}{N} \mat{\Sigma}^2\right)^{-1}\right) \mat{U}^\T,
\end{align}
where we have used $\mat{U} \mat{U}^\T=\mat{U}^\T \mat{U} = \mat{I}$.  Let us
define a diagonal matrix $\mat{J}$
%
%
\begin{align}
\label{eq:J}
J_{kk} &:= 1 - \left(1 + \frac{\left\langle A^2\right\rangle}{N} \sigma_k^2\right)^{-1} \nonumber \\
&= \frac{\sigma_{k}^2}{\sigma_{k}^2 + \frac{N}{\left\langle A^2\right\rangle}.}
\end{align}
With this definition the approximate detection statistic
\eqref{eqn:approxlambda} can be written as
%
%
\begin{eqnarray}
\Gamma &=& \mat{h}^\T \mat{U} \mat{J} \mat{U}^\T \mat{h},
\end{eqnarray}
which, in the notation of the previous section, is
%
%
\begin{eqnarray}
\label{eqn:statistic}
\Gamma &=& \sum_k \frac{\sigma_k^2}{\sigma_k^2 + N/{\langle A^2 \rangle}} (\mat{h}\cdot\mat{u_k})^2.
\end{eqnarray}

It is worth noting the limits of this expression for low and high 
amplitude signals (small and large values of $A$), 
%
%
\begin{subequations}
\begin{equation}
\label{eqn:Alowstatistic}
\lim_{A \to 0} \Gamma = \sum_k \sigma_k^2 (\mat{h}\cdot\mat{u_k})^2 ,
\end{equation}
\begin{equation}
\label{eqn:Ainfstatistic}
\lim_{A \to \infty} \Gamma = \sum_k (\mat{h}\cdot\mat{u_k})^2 .
\end{equation}
\end{subequations}
The $A \to 0$ limit is exactly the same result as the second order
approximation derived in the previous section.  In this limit the most
important filters (i.e., those with larger singular values) contribute more to
the composite detection statistic.  The $A \to \infty$ limit in
\eqref{eqn:Ainfstatistic} reduces the composite detection statistic to the sum
of squares of the orthogonal filter outputs, $(\mat{h}\cdot \mat{u_j})$.  This
detection statistic is equivalent to the excess power statistic obtained in
Anderson et.\ al for the detection of waveforms of known bandwidth and
duration~\cite[equation (2.10)]{Anderson:2000yy}.  Here, instead of projecting
the data onto a basis that spans a time-frequency tile, we project the data
onto a basis that span the space of CBC waveforms.  The essential difference
between \eqref{eqn:statistic} and the excess power statistic of Anderson et.\
al is that \eqref{eqn:statistic} folds in knowledge of the relative probability
that the waveforms we seek match any of the basis vectors individually, whereas
the target waveforms in~\cite{Anderson:2000yy} are assumed to match the basis
vectors of the time-frequency tile with equal probability.

\begin{figure}
\includegraphics{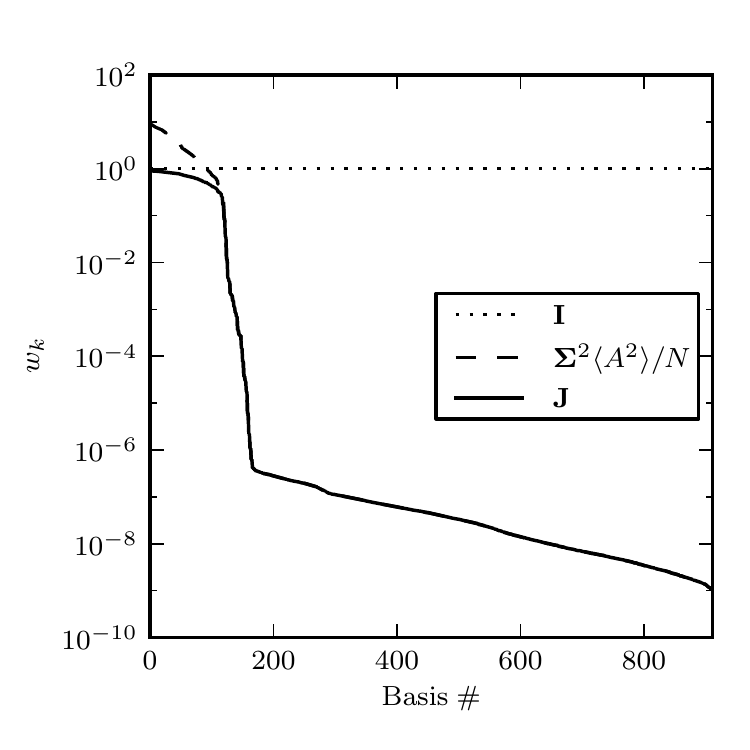}
\caption{The coefficients $w_k$ used to construct the composite detection
statistic as a function of basis vector number, ordered by their singular
value. The solid trace shows the coefficients we obtain from \eqref{eq:J}
assuming a signal amplitude  $A=20$. The dashed trace shows the singular values
squared, the choice of coefficients given in \eqref{eq:low_amp_likelihood}. The
dotted trace shows ``the excess power'' choice of coefficients, which are unity
for all basis vectors.}
\label{fig:coeff}
\end{figure}

Fig.~\ref{fig:coeff} presents an example of the components of \eqref{eq:J} for
the small and large amplitude limits, and for an amplitude of 20. These were
produced using the signal matrix described in Sec.~\ref{subsec:simulations}.
%
%


\section{Operating characteristics of the proposed composite detection statistics}
\label{sec:ROC}

In this section we explore the performance of the proposed detection statistic
\eqref{eqn:statistic}.  We begin by establishing the framework with which we
conduct simulations to produce \ac{ROC} curves. These indicate the probability
of detection versus the probability of false alarm.  We then present some
practical scenarios in which to understand these results.

We begin by assuming the detector data $\mat{h}$ has the form of
\eqref{eq:discrete_parameter_strain} modified to allow an ambiguous phase
%
%
\begin{equation}
\label{eq:discrete_parameter_strain_phase}
\mat{h} = \mat{n} + A \mat{s_{i(0)}} + B \mat{s_{i(\pi/2)}}.
\end{equation}
Under ideal situations the signal $\mat{s_i}$ could be recovered by a matched
filter with an \ac{SNR} of $\sqrt{A^2 + B^2}$.  According to~\cite{SVDpaper}
the signal $\mat{s_i}$ can be decomposed into orthogonal basis functions using
the singular value decomposition such that 
\begin{eqnarray}
\mat{s_{i(0)}} & = & \sum_{j} v^{ij}_{(0)} \sigma_j \mat{u_j} ,\\
\mat{s_{i(\pi/2)}} & = & \sum_{j} v^{ij}_{(\pi/2)} \sigma_j \mat{u_j} .
\end{eqnarray}
This leads to the following expression of the composite detection statistic
for \eqref{eq:discrete_parameter_strain_phase}
\begin{eqnarray}
\label{eq:trial}
\Gamma^2_i & = & \sum_k w_k (\mat{h} \cdot \mat{u_k})^2 \nonumber \\
& = & \sum_k w_k \left[A v^{ik}_{(0)} \sigma_k + B v^{ik}_{(\pi/2)} \sigma_k + n_k)\right]^2 \nonumber , \\
w_k & = & \frac{\sigma_k^2}{\sigma_k^2 + N/A^2} ,
\end{eqnarray}
where $n_k$ is a random number drawn from a unit variance Gaussian distribution.
We have made use of the fact that $(\mat{u_i}\cdot \mat{u}_j) = \delta_{ij}$.

In order to assess the operating characteristics of \eqref{eqn:statistic} we
simulate several instances of signals and several instances of noise.  We then
compare the number of noise trials that produce a value of
\eqref{eqn:statistic} greater than some threshold $\Gamma^*$ with the number
of signal, plus noise, trials that produce values above the same threshold.
This allows us to parameterize the detection probability versus false alarm
probability using the value of $\Gamma^*$.


\subsection{Simulations}
\label{subsec:simulations}
In this section we simulate the procedure described above.  Our goals are the
following. First we explore how the detection probability of
\eqref{eqn:statistic} varies of as a function of $A$ at a fixed false alarm
probability.  We verify that it peaks when the simulated signal amplitude is
equal to $A$.  We then compare the performance of \eqref{eqn:statistic} with
standard matched filtering results.  We find, to no surprise that
\eqref{eqn:statistic} alone performs worse.  However, we also find that by
using \eqref{eqn:statistic} to hierarchically reconstruct the physical template
SNR the same detection probability can be reached for sufficiently low false
alarm probability.

In order to conduct these tests, we apply the singular value decomposition to
\ac{BNS} waveforms with chirp masses $1.125 M_{\odot} \le M_c < 1.240
M_{\odot}$ and component masses $1 M_{\odot} \le m_1,m_2 < 3
M_{\odot}$~\cite{SVDpaper}. The number of templates required to hexagonally
cover this range in parameters using a minimal match of $96.8\%$ is $M=456$,
which implies a total number of single-phase filters $N=912$.  These
non-spinning waveforms were produced to 3.5PN order\cite{LAL}, sampled at 2048
Hz, up to the Nyquist frequency of 1024 Hz.  The last 10 seconds of each
waveform, whitened with the initial LIGO amplitude spectral density, were used
to construct the matrix of signals $\mat{S}$. 

\begin{figure}
\includegraphics{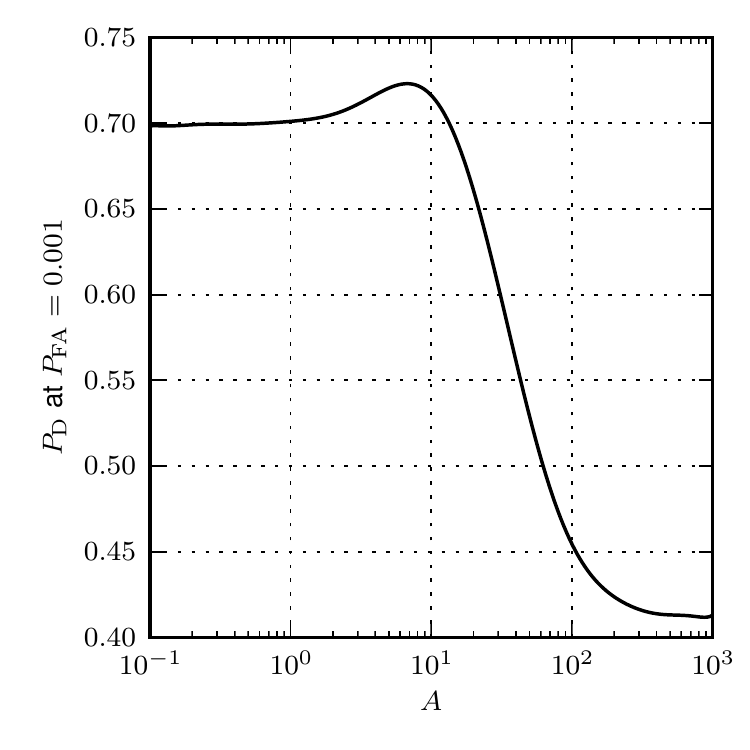}
\caption{Detection probability at a false alarm probability of $10^{-3}$ as a
function of $A$.  We simulated $\sim 1.8\times10^7$ signals at SNR 7 for 300
values of $A$ in the composite detection statistic defined by
\eqref{eqn:statistic}.  As we expect the detection probability peaks when $A=7$,
the amplitude of the signal.  It is worth noting that in this case the low
amplitude limit given by \eqref{eqn:Alowstatistic} provides a similar detection
probability.  However the large signal limit given by \eqref{eqn:Ainfstatistic}
is considerably worse.
\label{fig:A}}
\end{figure}

Using this framework we first test how the detection probability varies with
the choice of $A$ in \eqref{eqn:statistic}.  We simulate $\sim 1.8\times10^7$
signals with SNR $7$ and evaluate \eqref{eqn:statistic} as a function of $A$.
Likewise we evaluate \eqref{eqn:statistic} for just noise.  The results of the
detection probability at a false-alarm-probability of $10^{-3}$ are shown in
Fig.~\ref{fig:A}.  We find that the peak of the detection probability occurs
when $A$ equals the amplitude of the signal, 7.  We note that for SNR 7 signals
the low amplitude limit expression \eqref{eqn:Alowstatistic} performs nearly as
well (within a few percent). However, the high amplitude limit is considerably
worse (by almost a factor of two).  SNR 7 was chosen to accentuate the
dependency of \eqref{eqn:statistic} on $A$.  SNR 7 is actually higher than the
typical SNR threshold one would place in a gravitational-wave search.  

\begin{figure}
\includegraphics{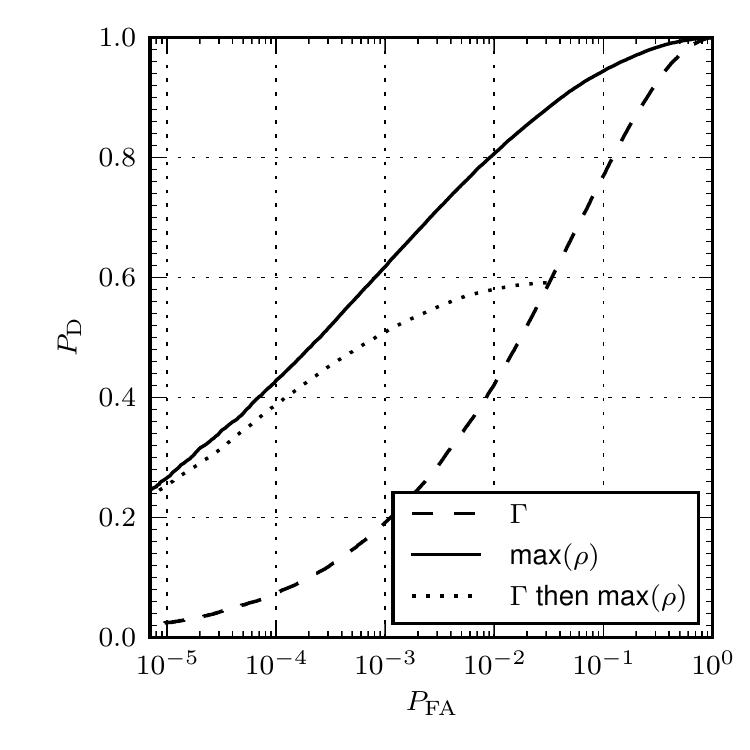}
\caption{Receiver Operator Characteristic curves associated with different
detection statistics. The solid-black trace shows the performance of choosing
the maximum likelihood filter across the bank.  The dashed line shows the
performance of the composite detection statistic.  The dotted line shows
the performance of choosing max filter across the bank after conditionally
thresholding on the composite detection statistic.}
\label{fig:ROC}
\end{figure}

In order to test the efficiency of the composite detection statistic versus the
traditional matched filter approach we again generated $\sim 1.8\times10^7$
instances of \eqref{eq:trial} for signal and signal plus noise.  This time we
chose a lower, more realistic, signal amplitude of $5$.  The signals had
uniform distributions in the template bank and in phase angle.  We compare this
with the standard result arising from maximizing the SNR across the bank and
over phase angle 
\begin{equation}
\label{eq:max_likelihood}
\rho_{\mathrm{max}} := \mathrm{max}_i\left[ (\mat{h} \cdot \mat{s_{i(0)}})^2 + (\mat{h} \cdot \mat{s_{i(\pi/2)}})^2 \right].
\end{equation}
The result of the two procedures is shown in Fig.~\ref{fig:ROC}.  As expected,
the composite detection statistic performs worse than explicit reconstruction
of the template parameters.  However, when the two methods are combined, it is
possible to reach the same detection probability with asymptotically low false
alarm probability.

Fig.~\ref{fig:ROC} contains three curves, showing the detection probability
$P_{\mathrm{D}}$ versus false alarm probability $P_{\mathrm{FA}}$ when there is
a signal with amplitude $A=5$ in the data.  The solid-black line is found by
choosing the maximum \ac{SNR} across the filter bank as defined in
\eqref{eq:max_likelihood}.  Maximization over the filter output is commonly
done in gravitational-wave searches.  The dashed-black line is the result of
the composite detection statistic \eqref{eqn:statistic} with the choice of
$A=5$.  The dotted-black line is the result of first thresholding on the
composite detection statistic and then conditionally maximizing the \ac{SNR}
over the bank.  This procedure produces roughly the same detection probability
for a false alarm probability of $10^{-5}$ but allows one to do the full
maximization for only 3\% of the filtered data.  

\subsection{Use example}

Our simulations indicate that only $\sim 3\%$ of the data needs to have the
physical template parameters reconstructed in order to have a similar
detection probability as the maximum likelihood method at a false alarm
probability of $10^{-5}$.  This section provides an example of what this means
for a realistic gravitational-wave search.

Advanced gravitational-wave detectors should be able to analyze and locate
compact binary sources at the moment they merge.  Prompt electromagnetic
followup could confirm a gravitational-wave detection and low-latency searches
will be critical to maximize the number of simultaneously observed signals.
However, low-latency gravitational-wave searches will be computationally
costly.  The reduced filter set proposed by~\cite{SVDpaper} could lower the
computational cost substantially, helping to enable near real-time searches.
Additionally, in this work, we have shown that it is possible to reduce the
number of physical parameter reconstructions by 97\% and maintain similar
detection efficiencies.  If these methods were to be used, it would be
necessary to understand what the result in Fig.~\ref{fig:ROC} implies.

We now consider how the results of Fig.~\eqref{fig:ROC} may apply to a
low-latency gravitational-wave search and answer whether or not a false-alarm
probability of $10^{-5}$ is a useful operating point.  Consider the joint false
alarm probability for N independent gravitational-wave detectors in coincidence
\begin{equation} 
P_{\mathrm{FA}} = \mathcal{C}^{N-1} \times \prod_i^N P_{\mathrm{FA},i} , 
\end{equation} where $P_{\mathrm{FA},i}$ is the false alarm
probability for the $i$th detector and $\mathcal{C}$ is the coincidence trials
factor.  In order to understand the double coincidence, limiting false alarm
rate from the single detector false alarm probability a few pieces of
information are needed.  The first is the number of independent trials per
second obtained by filtering the data.  We will take this to be the frequency
of the minimum point of the noise curve $\sim 150$ Hz.  If one allows $\sim 30$
ms to define coincidence this corresponds to an additional 5 samples for
coincidence trials at 150 Hz.  Therefore the false alarm rate of double
coincidence corresponding to a $10^{-5} $ false alarm probability in a single
detector is $150 \mathrm{Hz} \times 5 \times10^{-10} = 7.5\times10^{-8}
\mathrm{Hz} = 2.4 \mathrm{yr}^{-1}$.  This is well above the false rate that
would be required for a detection candidate.  Therefore we conclude that this
procedure should not impact the detectability of near threshold signals.


\section{Conclusions}

We have presented a study of compact-binary gravitational-wave detection that
precedes parameter estimation.  This could allow more computationally efficient
algorithms to be run in near real time that determine whether a signal is
present before attempting to measure it's parameters.  Our study shows that it
should be possible to reconstruct the physical parameters for only
$\mathcal{O}[1\%]$ of the data while not impacting the sensitivity of a compact
binary search.


\section{Acknowledgements}

The authors would like to acknowledge the support of the LIGO Lab, NSF grants
PHY-0653653 and PHY-0601459, and the David and Barbara Groce Fund at Caltech.
LIGO was constructed by the California Institute of Technology and
Massachusetts Institute of Technology with funding from the National Science
Foundation and operates under cooperative agreement PHY-0757058.  Research at
Perimeter Institute is supported through Industry Canada and by the Province of
Ontario through the Ministry of Research \& Innovation. KC was supported by the
National Science and Engineering Research Council, Canada.  DK was supported in
part from the Max Planck Gesellschaft.  CH would like to thank Patrick Brady
and DK would like to thank Christopher Messenger and Reinhard Prix for many
fruitful discussions concerning this work.  The authors would also like to
thank Alan Weinstein for useful comments on this manuscript. This work has LIGO
document number {LIGO-P1000038-v1} and Perimeter Institute report number
{pi-other-203}.


\bibliography{references}

\end{document}